\documentclass[12pt]{article}
\usepackage{graphicx}
\usepackage{rotating}

\newcommand{\um}{$\mu$m}
\newcommand{\kms}{km s$^{-1}$}
\newcommand{\lsun}{L$_{\odot}$}

\begin{document}
\Large
{\bf The ISO-LWS far-infrared spectra of pre-Main Sequence
objects.\footnote{Based on observations with Infrared Space Observatory, an ESA
project with instruments funded by ESA Member States (especially the PI
countries: France, Germany, the Netherlands and the United Kingdom) and with the
participation of ISAS and NASA}}
\vspace{1cm}

\large
Milena Benedettini, Stefano Pezzuto, Luigi Spinoglio, Paolo Saraceno, Anna Maria
Di Giorgio
\vspace{0.5cm}

CNR -- Istituto di Fisica dello Spazio Interplanetario, Area di Ricerca
di Tor Vergata, via del Fosso del Cavaliere 100, 00133, Roma, Italy

\vspace{4cm}

Running title: The ISO-LWS FIR spectra of PMS objects.

\vspace{5cm}

To be published in the book ``Recent Research Developments in Molecular and
Cellular Astronomy \& Astrophysics''

\newpage

\normalsize

\section*{ABSTRACT}

The Long Wavelength Spectrometer (LWS) on board the Infrared Space Observatory
(ISO) satellite has allowed for the first time to investigate the whole far
infrared (FIR) part of the spectrum (from 43 to 197 \um), inaccessible from the
ground. The FIR spectrum is crucial for the study of the early phases of
protostellar evolution since the peak of the continuum emission of the
protostars falls around 100 \um\, and many molecular species emit in this range.

In this paper we review the main results of the spectroscopic observations
carried out with the LWS on a sample of pre-Main Sequence objects in different
evolutionary stages. The FIR spectra are characterized by a strong continuum
with superimposed atomic and molecular emission lines. The fine structure lines
emitted by the OI and CII atoms are present in all the objects of our sample
while emission from molecular species (CO, H$_2$O and OH) dominate the spectrum
of the younger and more embedded objects; lines of highly ionized atoms (OIII,
NIII, NII) are present only in high luminosity sources. We will show how both
the FIR continuum and the emission lines are powerful indicators of the
evolutionary status of young embedded objects.

The analysis of the LWS spectra has allowed to study only the average properties
of the young stellar objects, due to the large field of view
($\sim$ 80 arcsec), the limited sensitivity and the poor spectral resolution
($\sim$ 200) of the instrument in its grating mode. The higher spatial and
spectral resolutions of the instruments on board the Herschel satellite will
allow to address in deeper detail the scientific problems highlighted by
the ISO results.

\section{INTRODUCTION}

During the early phases of the star formation process the spectral energy
distribution (SED) of a protostar has the shape of a cold blackbody (few tens of
K) modified by the emissivity of the dust of the molecular
cloud where the object is forming. The central object, even when already formed,
is not visible at any wavelength since the surrounding dust envelope completely
obscures the inner region. As the evolution goes on, the optical depth of the
dust decreases making the central object visible in the near infrared until it
appears in the optical and its spectrum looks like that of a normal star with an
infrared excess due to the residual circumstellar material. At the typical
temperatures of pre-Main Sequence (PMS) objects (T=30-80 K) the dust emission
peaks in the far infrared (FIR) between 50-200 $\mu$m. The warm circumstellar
envelope is predicted to emit a rich FIR spectrum \cite{cec96} where the
prominent features are the water and carbon monoxide rotational lines and the
atomic oxygen fine structure lines.

The early stage of the star formation process is also characterized by the
presence of energetic mass ejections which interact with the surrounding
molecular cloud giving rise to bipolar outflows. The molecular outflow is
thought to be the mechanism through which the accreting central object loses the
exceeding angular momentum and therefore it plays an important role in
stopping the gravitational collapse, thus setting the final mass of the star.
Molecular outflows display a large range of excitation conditions, due to the
complex way in which the interaction between stellar winds/jets and ambient
medium is taking place. Ground based observations of molecular emission lines at
mm wavelengths, trace the large scale material at excitation temperatures of
about 10-20 K and vibrational lines of molecular Hydrogen at NIR wavelengths
probe the highly excited gas (at T$_{ex} \sim$ 10$^4$ K) along the flow axis.
However the presence of gas excited at intermediate temperature, i.e. 100-2000
K, can only be traced by FIR space borne observations.

The evolutionary status of the PMS objects cannot be
derived by directly observing the stellar radiation, which is only
visible during the late phases of star formation, but can be estimated only
indirectly by observing the continuum, atomic and molecular emission of the
circumstellar envelope or/and of the outflow. Unfortunately, most of the
emission falls in the FIR and it is inaccessible from the ground.
So far, the methods commonly used to classify the PMS objects have been mainly
based on the characterization of the shape of the SED at near and mid infrared
wavelengths through the
value of the spectral index $\alpha$\footnote{The spectral index $\alpha$ is
defined as: $\alpha=\frac{dLog(\nu F_{\nu})}{d\nu}$}. Lada \& Wilking \cite{lw}
first classified low-mass protostellar objects considering
three different classes defined by the value of $\alpha$ in the range 2--10
\um\, (then extended to 25 \um\, with the IRAS mission). This classification was
followed by the work of Adams, Lada \& Shu \cite{als} who associated to each
class a different evolutionary stage. Successively Andr\`e, Ward-Thompson \&
Barsony \cite{AWB1} introduced a new class of objects still younger than the
previous classes. The net result of all these works is the classification of
low-mass pre-Main Sequence (PMS) sources in four different classes (see Fig.
\ref{classes}), proceeding from the youngest to the oldest they are:

\begin{figure}
  \includegraphics[width=12cm]{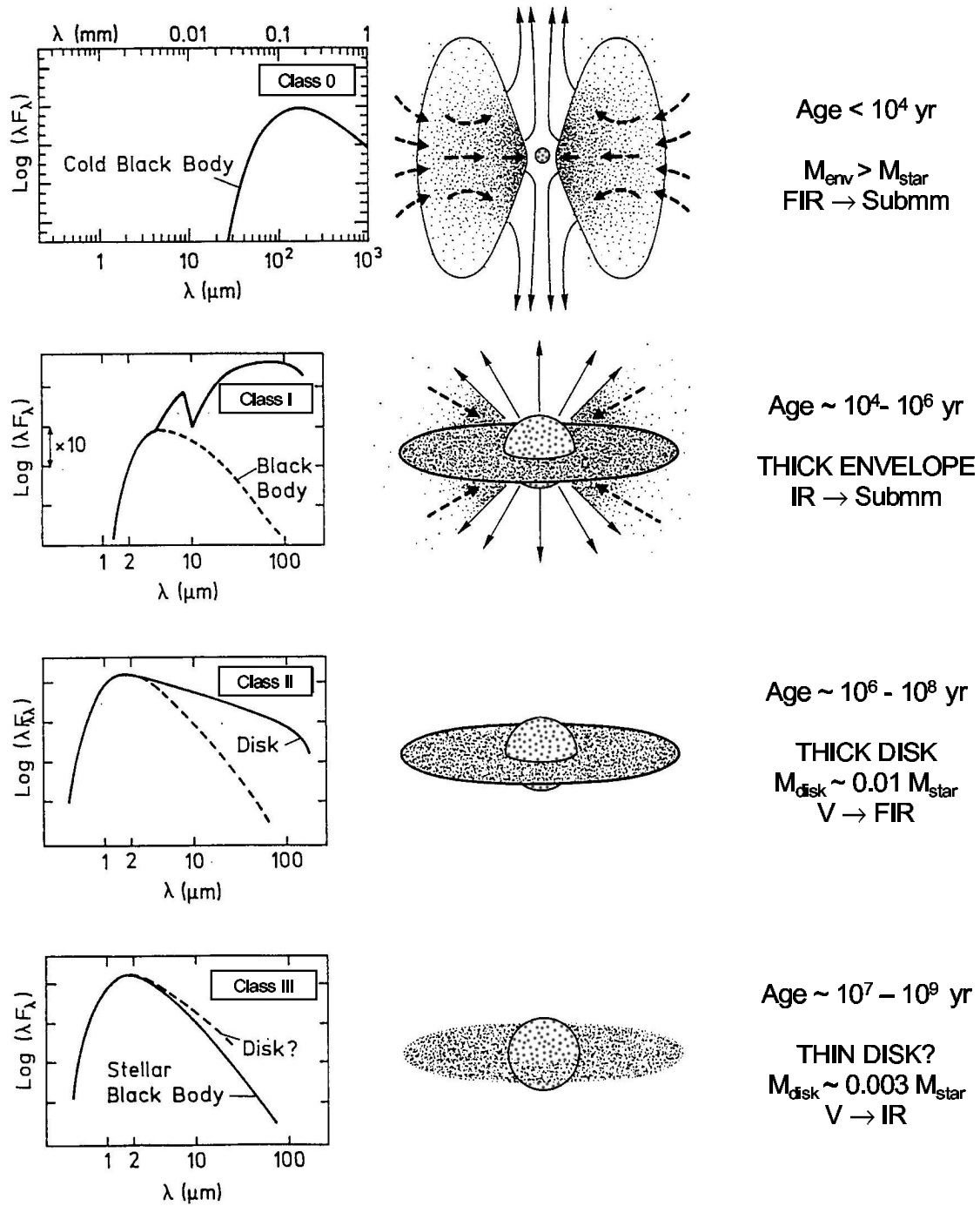}
  \caption{A scheme of the classification of the low-mass PMS objects, adapted
from the original figure of \cite{andre94}.}
  \label{classes}
\end{figure}

\begin{itemize}
\item Class 0: since they are visible only at $\lambda>$ 10 \um, the spectral
index $\alpha$ cannot be derived. They are identified by the ratio between the
submillimeter ($\lambda>$350\um) and the bolometric luminosity
L$_{\rm submm}$/L$_{\rm bol}>$5$\times$10$^{-3}$. They are newly formed
protostars in
the main accretion phase, when the mass of the central object is still less than
that of the circumstellar material. The central source is completely obscured by
the envelope. The spectrum is that of gray body with temperature $\sim$
25--35 K. Strong and collimated outflows are associated with these objects which
have ages $<$10$^4$ yr.
\item Class I: with $\alpha<$ 0 between 2 and 25 \um. The spectrum steeply
increases in the mid-infrared due to the optically thick envelope. They
represent protostars in the final phase of accretion when the stellar winds are
dispersing the circumstellar envelope but usually they are not yet visible in
the optical. The typical age is between 10$^4$ to 10$^6$ yr.
\item Class II: with 0 $<\alpha<$ 2 between 2 and 25 \um. The spectrum appears
in the optical and has an excess in the infrared due to the circumstellar
envelope or/and to an optically thick disk. They are identified with the
low-mass classical T Tauri stars and the intermediate-mass Herbig Ae/Be stars.
The typical age is between 10$^6$ to 10$^8$ yr.
\item Class III: with $\alpha>$ 2 between 2 and 25 \um. The spectrum is that of
a black body with a weak infrared excess due to the residual circumstellar
material. These objects are approaching the Main Sequence and do not have an
accretion disk or have only an optically thin disk. They are identified with the
Weak T Tauri stars (WTT) and have typical age of 10$^7$ -- 10$^9$ yr.
\end{itemize}

With the advent of the infrared satellites: IRAS first and ISO later, the FIR
observations have became available, boosting the study of the early phase of
star formation. In particular the Long Wavelength Spectrometer (LWS) on board
the ISO has allowed for the first time to measure the FIR spectrum of the
protostars. In this paper we review the main results of the observations carried
out with the LWS on a sample of PMS objects in different evolutionary stages and
we will show how both the FIR continuum and emission lines are powerful
indicators of the evolution.

\section{THE SAMPLE OF PRE-MAIN SEQUENCE OBJECTS OBSERVED BY THE LWS}

We consider the sample of PMS objects observed in the LWS star formation core
program\footnote{This program has been performed as part of the Guaranteed Time
of the LWS Consortium.}. The aim of the program was to study the first stages of
stellar evolution from the beginning of the accretion process, when the
protostar is deeply embedded in the parent molecular cloud and most of the
energy is given by the conversion of gravitational energy into radiation, till
the end of the accretion, when the envelope is disrupted by the stellar winds
and the protostar becomes optically visible. Since all young stars experience
outflow phenomena during the earliest phases of their evolution, the study of
the outflow and its interaction with the interstellar medium has been considered
essential for the purposes of the program. The sample was composed by objects
belonging to Class 0, Class I and Class II, selected both for having a strong
submillimeter continuum and for being associated with a molecular outflows.

\begin{table*}
\centering
\hbox{
\footnotesize
  \rotcaption{The sample of PMS objects observed. Sources have been grouped
according  to their known evolutionary status. An asterisk flags objects
classified as FU  Orionis. The reference where the LWS spectrum is discussed in
detail is given.}
  \label{list}
\begin{sideways}
    \begin{tabular}[b]{lcccclcccc}
      \hline
&IRAS ID&\multicolumn{2}{c}{coordinates}&Ref.&
&IRAS ID&\multicolumn{2}{c}{coordinates}&Ref.\\
  &&\multicolumn{1}{c}{$\alpha$(J2000)}&\multicolumn{1}{c}{$\delta$(J2000)}&&
  &&\multicolumn{1}{c}{$\alpha$(J2000)}&\multicolumn{1}{c}{$\delta$(J2000)}\\
\bf Class 0      &&&&&\bf Bright Class~I\\
L 1448 IRS3      &&03 25 36.3&30 45 15.0&\cite{nisi2000}&
NGC 281 W    &00494+5617&00 52 23.7&56 33 45.9&\cite{monti}\\
L 1448 mm        &&03 25 38.8&30 44 05.0&\cite{nisi1999}&
AFGL 437     &03035+5819&03 07 23.7&58 30 52.0&\cite{monti}\\
NGC 1333 IRAS 2  &03258+3104&03 28 55.4&31 14
35.0&\cite{cec1999},\cite{gian2001}&
AFGL 490     &03236+5836&03 27 38.8&58 47 00.4 &\cite{monti}\\
NGC 1333 IRAS 6  &&03 29 02.8&31 20 30.8&\cite{nisi2002}&
NGC 2024 IRS2    &&05 41 45.8&-01 54 29.3&\cite{gian2000}\\
NGC 1333 IRAS 4  &&03 29 11.1&31 13 20.3&\cite{cec1999},\cite{gian2001}&
NGC 6334 I       &&17 20 53.4&-35 47 50.9&\cite{monti}\\
IRAS03282+3035&03282+3035  &03 31 20.3&30 45 24.8&\cite{gian2001}&
W 28 A2      &&18 00 30.3&-24 03 12.3&\cite{monti}\\
L1641 VLA 1      &&05 36 22.8&-06 46 08.6&\cite{gian2001}&
M 8 E        &&18 04 52.7&-24 26 36.2&\cite{monti}\\
NGC 2024 FIR 3   &&05 41 42.9&-01 54 23.6&\cite{gian2000}&
GGD 27 IRS   &18162-2048&18 19 12.1&-20 47 31.7&\cite{moli01}\\
NGC 2024 FIR 5   &&05 41 44.2&-01 55 37.7&\cite{gian2000}&
S 87 IRS 1   &19442+2428&19 46 20.1&24 35 29.6&\cite{monti}\\
HH 25 mm         &&05 46 06.6&-00 13 24.6&\cite{bene2000}&
NGC7538 IRS5     &&23 13 31.4& 61 29 50.7&\cite{monti}\\
HH 24 mm         &&05 46 08.6&-00 10 40.9&\cite{bene2000}&
NGC7538 IRS1     &&23 13 49.1& 61 28 9.7&\cite{monti}\\
VLA 1623         &&16 26 26.3&-24 24 29.7&  &
NGC7538 IRS9     &&23 13 58.1& 61 27 18.9&\cite{monti}\\
IRAS16293$-$2422&16293$-$2422&16 32 22.7&-24 28
33.2&\cite{gian2001},\cite{cec1998} &
\bf Herbig Ae/Be\\
L 483        &18148$-$0440&18 17 29.8&-04 39 38.6&\cite{gian2001}&
LK H$\alpha$ 198&00087+5833&00 11 26.0&58 49
29.5&\cite{loren1999}\\
Serpens FIRS 1&18273+0113&18 29 50.3&01 15 18.3&\cite{gian2001}&
V376 Cas     &&00 11 26.7&58 50 04.5&\cite{loren1999}\\
L 723            &19156+1906&19 17 53.2&19 12 14.6&\cite{gian2001}&
*Z CMa       &07013$-$1128&07 03 43.1&-1133 06.0&\cite{loren2000}\\
B 335            &19345+0727&19 37 00.8&07 34 10.8&\cite{nisi1999a}&
HD 97048&11066$-$7722&11 08 04.6&77 39 16.9&\cite{loren1999}\\
\bf Class~I      &&&             &&
DK Cha  &12496$-$7650&12 53 16.4&77 07 01.9&\cite{loren1999},\cite{gian1999}\\
SVS 13           &03259+3105&03 29 03.8&31 16 02.9&\cite{mol2000}&
CoD $-39^\circ$ 8581&13547$-$3944&13 57 41.2&-39 58 43.6&\\
L 1551 IRS 5     &04287+1801&04 31 34.1&18 08 04.8&\cite{white2000}&
CoD $-42^\circ$ 11721&16555$-$4237&16 59 06.8&-42 42
07.6&\cite{loren1999},\cite{bene1998}\\
L1641 N          &05338$-$0624&05 36 19.0&-06 22 13.2&&
MWC 297&18250$-$0351&18 27 39.5&-03.49.52.1&\cite{loren1999},\cite{bene2001}\\
HH26 IRS     &&05 46 03.9&-00 14 52.4&\cite{bene2000}&
R CrA        &18585$-$3701&19 01 53.6&-36 57
08.7&\cite{loren1999},\cite{gian1999}\\
HH 46 IRS        &08242$-$5050&08 25 43.8&-51 00 35.6&\cite{nisi2002}&
T CrA        &&19 01 58.8&-36 57 49.4&\\
WL 16            &&16 27 02.0&-24 37 25.9&\cite{nisi2002}&
PV Cep&20453+6746&20 45 54.1&07 57 39.1&\cite{loren1999}\\
Elias 29         &&16 27 09.3&-24 37 18.4&\cite{nisi2002}&
V645 Cyg&21381+5000&21 39 58.2&50 14 21.8&\cite{loren1999}\\
$\rho$ Oph IRS 43&&16 27 27.0&-24 40 49.0&\cite{nisi2002}&
LK H$\alpha$ 234&21418+6552&21 43 06.6&66 06
54.4&\cite{loren1999},\cite{gian1999}\\
$\rho$ Oph IRS 44&&16 27 28.0&-24 39 32.9&\cite{nisi2002}&
MWC 1080     &23152+6034&23 17 25.8&60 50
43.5&\cite{loren1999},\cite{bene1998}\\
$^*$Re 13    &16289$-$4449&16 32 27.1&-44 55 36.1&\cite{loren2000}&
\bf T Tauri\\
$^*$HH 57 IRS    &16289$-$4449&16 32 32.0&-44 55 28.9&\cite{loren2000}&
*RNO 1B      &&00 36 46.2&63 29 54.2&\cite{loren2000}\\
L 379 IRS 3      &18265$-$1517&18 29 24.7&-15 15 48.6&&
T Tau&04190+1924&04 21 59.4&19 32 06.3&\cite{spino}\\
IC 1396 N        &&21 40 42.3&58 16 09.8&\cite{sara1996}&
DG Tau       &04240+2559&04 27 04.7&26 06 16.5&\\
NGC 7129 FIRS 2  &&21 43 00.8&66 03 37.1&&
HL Tau&04287+1807&04 31 38.4&18 13 58.5&\\
L 1206           &22272+6358&22 28 51.5&64 14 32.8& &
SR 9         &&16 27 403&-24 22 02.2&\\
&&&&&
*Elias 1-12&21454+4718&21 42 20.6&47 32 04.9&\cite{loren2000}\\
\hline
      \end{tabular}
\end{sideways}
}
\end{table*}

In Table \ref{list} we list the objects of the sample together with their
coordinates and the reference of the papers where the LWS spectrum is discussed
in detail. Our sample consists of 64 PMS objects that can be divided in five
groups: Class 0, Class I, Luminous Class I (characterized by having high
infrared luminosity, i.e. L$_{\rm IRAS}>$10$^4$ \lsun), Herbig Ae/Be
(intermediate-mass Class II) and T Tauri (low-mass Class II). For those
sources that have extended outflows, additional observations mapping the lobes
of the outflow have been obtained. The observations have been carried out with
LWS in its full range grating mode which provides spectra between 43 and 197
\um\, with a resolution $\lambda/\Delta\lambda\sim 200$ and an instrumental beam
size of $\sim$80 arcsec (for a detailed description of the instrument see
\cite{clegg}). Data have been reduced with the same standard method using
the official software provided by the LWS and SWS consortiums.

\section{THE EVOLUTION THROUGH THE FIR CONTINUUM EMISSION}

The LWS wavelength range is the most suitable to derive the evolutionary status
of PMS objects because it includes the peak of the dust emission. To this aim a
two colour\footnote{ The colour $[\lambda_i -\lambda_j]$ is defined as:
$[\lambda_i -\lambda_j]=\log\frac{F_j}{F_i}$ where $F_{i,j}$ is the flux
densities in Jy, averaged over a band, at the wavelengths $\lambda_{i,j}$.}
diagram based on LWS data has been used as a diagnostic tool \cite{pez}. A
similar diagram was introduced already for the IRAS data to infer the
evolutionary phase of PMS objects (e.g. \cite{beich}, \cite{berri}), however the
LWS extends the FIR spectral coverage of IRAS by a factor $\sim 2$ in the region
where the emission of blackbodies of 15 -- 30 K peaks. Compared to the previous
evolutionary diagnostic tools, like the spectral index $\alpha$ \cite{als}, the
bolometric temperature \cite{ml}, \cite{chen}, and the $L_{\rm bol}$ vs. $F_{\rm
1.3mm}$ diagram \cite{paolo}, this method can be applied to PMS objects in all
evolutionary phases (Class 0, Class I and Class II objects), and does not
require the knowledge of either the bolometric luminosity or the complete SED.

\begin{figure}
  \includegraphics[width=15cm]{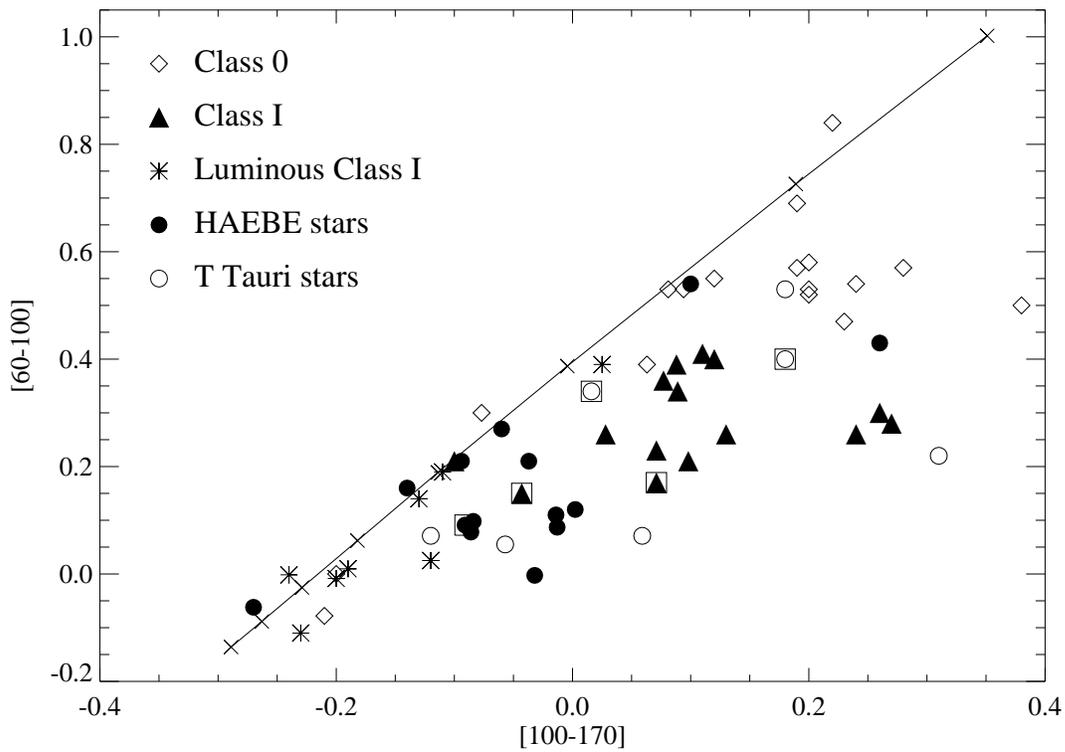}
  \caption{Two colour diagram for the sample of PMS objects from \cite{pez}.
Boxed
   symbols identify FU Orionis objects. The error, not displayed, on each point
   is 0.26 in both colours. The solid line connects the  blackbody colours at
temperatures (marked by crosses): 25\,K (upper right corner), 30, 40, 50, 60,
70, 80 and 90\,K (lower left corner). For a discussion on individual sources
with an anomalous position see \cite{pez}.}
  \label{fig1}
\end{figure}

For the LWS data Pezzuto et al. \cite{pez} introduced ``photometric'' bands at
60 and 100\um\, to allow a comparison with the IRAS data, and at 170\um. The
flux level was derived averaging the spectra over 1 \um\, around the three
wavelengths and the flux uncertainty is 30\%, which turns into an error on each
colour of 0.26. The [60--100] vs. [100--170] diagram is shown in
Figure~\ref{fig1}, where for comparison it is also reported the colour of a
blackbody at various temperatures (marked by crosses and connected by the solid
line) ranging from 25~K (top right) to 90~K (bottom left).

Each PMS class occupies a well defined region of the diagram with a clear trend
of increasing colour temperature $T_{\rm c}$ with the age of the sources. On
average, Class 0 objects have $T_{\rm c}\sim 30\,$K, Class I $T_{\rm c}\sim$
35-40 K and Class II $T_{\rm c}\sim$ 50 K. Luminous Class I objects show the
highest temperatures with $T_{\rm c}\sim$ 50-80 K. This trend can be
qualitatively explained by considering that at these wavelengths $T_{\rm c}$
depends only on the temperature stratification of the circumstellar matter. In
the first stage of the evolution we can see only the outer region of the
envelope at cold temperatures. The optical depth of the envelope decreases with
the age so that we can look deeper into the inner, and warmer, regions of the
envelope and, consequently at higher $T_{\rm c}$.

The average colours are systematically under the curve of the blackbody colours,
this is probably due to the combination of both the background contamination and
the confusion of the observed sources in the large LWS beam ($\sim$ 80 arcsec).
The low spatial
resolution of the instrument is indeed the main limitation of this result.
In this respect this work will be greatly improved by future space missions
which will be able to observe the same regions with a higher spatial resolution
e.g. PACS, the FIR camera onboard the Herschel satellite.

\section{THE FIR EMISSION LINES SPECTRUM}
\subsection{CLASS 0, I AND II}

\begin{figure}
  \includegraphics[width=15cm]{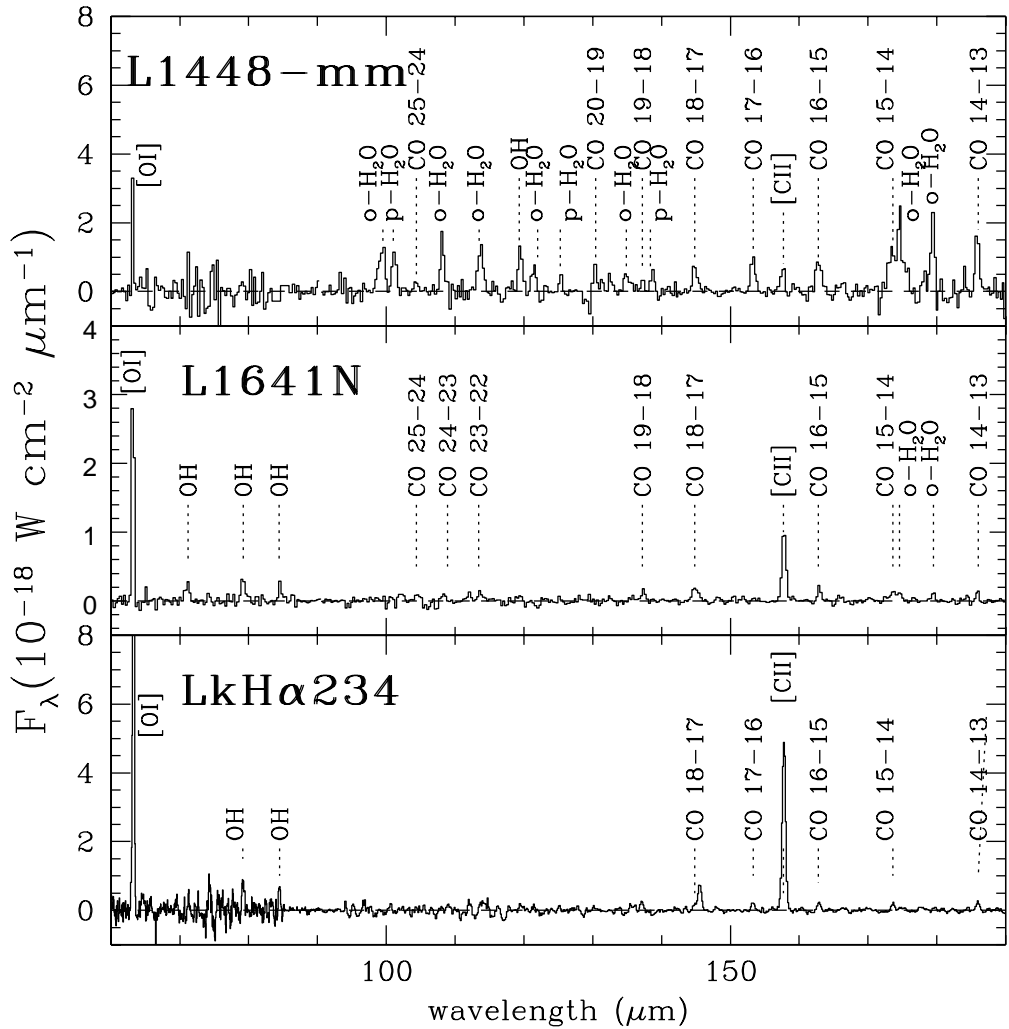}
  \caption{Continuum subtracted LWS spectra of the Class 0 L1448-mm (top panel),
the Class I L1641N (middle panel) and the Class II LkH$\alpha$234 (bottom
panel).}
  \label{spectra}
\end{figure}

In Figure \ref{spectra} we present the LWS spectrum of three objects considered
as a prototype of their evolutionary class. This figure indicates that the FIR
spectra of PMS sources are correlated to the evolutionary status of the source
itself, the richer spectra being that of the youngest Class 0 objects. For what
concern the molecular emission, Class 0 sources mainly emit through pure
rotational lines of high-J (J$_{\rm up}>$14) CO and H$_2$O molecules, OH lines
are also observed in about 40\% of the sample. In Class I sources the molecular
emission is less intense with respect to Class 0: even if the high-J CO lines
are still present in most of the sources, the emission from water is rarely
detected. Finally among the Class II objects the molecular (CO and OH) emission
is present only in 3 out of 14 Herbig AeBe stars.

\begin{figure}
  \includegraphics[width=13cm]{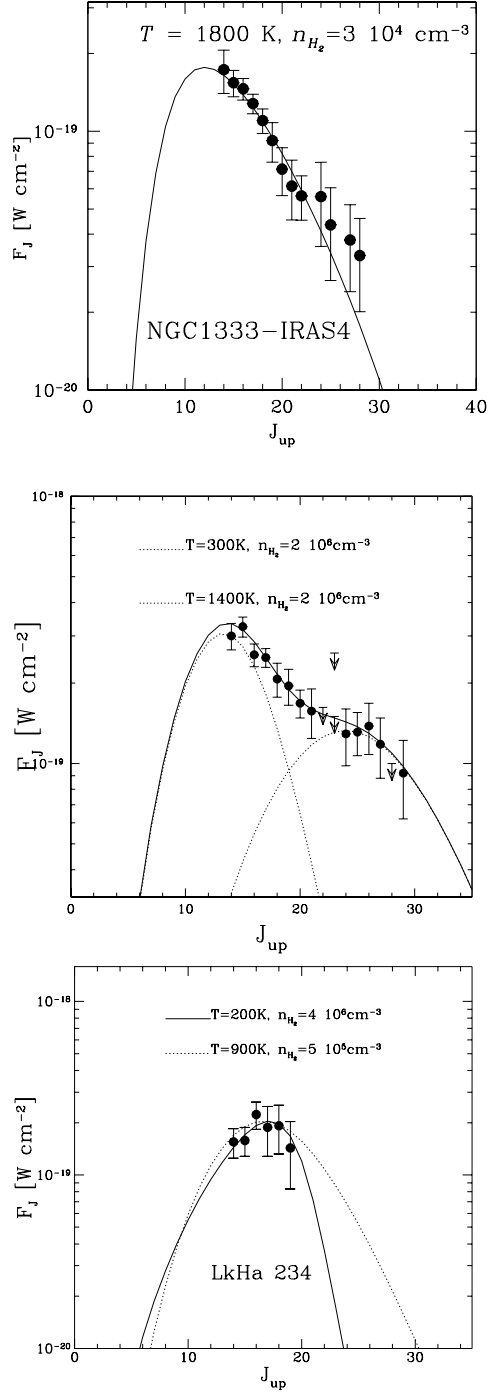}
  \caption{Observed CO line fluxes as function of the rotational quantum number
$J_{\rm up}$. The lines are the LVG model fit to the data, whose
parameters are indicated. The figures are taken from Giannini et
al. \cite{gian2001} for NGC1333 IRAS4 (top panel), from Saraceno
et al. \cite{sara2001} for IC 1396 N (middle panel) and from
Giannini et al. \cite{gian1999} for LkH$\alpha$ 234 (bottom
panel) }
  \label{fitco}
\end{figure}

The molecular lines observed on-source can originate either from the envelope
surrounding the source or in the shock associated with its powerful outflow. The
relatively low spatial and spectral resolution of the LWS does not allow to
discriminate between the two hypotheses about the origin of the these lines.
Ceccarelli et al. \cite{cec1999} are in favour of the first hypothesis, in
particular for the water lines, since they found that in five sources the water
emission correlates with the millimeter continuum emission that is a direct
measurement of the envelope mass, whereas it does not correlate with the SiO
emission observed towards the same sources, that is an indicator of the shock
activity. Similar woks show how the water lines observed towards two Class 0
objects (i.e. IRAS16293-2422 \cite{cec2000} and NGC1333-IRAS4 \cite{maret}) can
be reproduced with a model of a collapsing envelope. However these wokrs
do not account for the other atomic (OI) and molecular (CO) lines detected on
the same source and they do not explain the fact that similar flux levels
of the observed FIR lines are detected both on the central source and on the
lobes of the associated outflow. This observational constraint would favour the
second and most widely accepted interpretation that the observed molecular lines
are excited by the shocks associated with the outflow. In fact, a combination of
a dissociative J shock occurring at the apex of the stellar jet/wind and a
non-dissociative C shock originating in the wings of a bow-like flow can account
for most of the observed emission (e.g.
\cite{gian2001},\cite{nisi2002},\cite{bene2000} and references therein). The
emission lines from CO, H$_2$O and OH are usually simultaneously fitted by Large
Velocity Gradient (LVG) models indicating that they arise from small regions,
with a typical angular size of 10$^{-9}$ sr, of warm (200$<T$$<$2000 K) and
dense (10$^4<$n(H$_2$)$<10^7$ cm$^{-3}$) gas compressed by the passage of the
shock. In Figure \ref{fitco} the LVG fit of the CO lines is shown for three
sources: NGC1333 IRAS4 (Class 0), IC 1396 N (Class I) and LkH$\alpha$ 234 (Herbig Ae/Be).
For LkH$\alpha$ 234 as well as for most of the sources of the sample, a range of possible
models fits the data: the spread in the resulting parameters is relatively high because of
the high errors in the measured fluxes and the small number of lines detected. In IC 1396 N
the higher J$_{\rm up}$ lines cannot be fitted by the same parameters as the other lines,
suggesting the present of a second component at higher temperature.

Observations with the Submillimeter Wave Astronomy Satellite (SWAS) of the
o-H$_2$O ground level transition at 538\um\, are
available for four outflows sources of the ISO sample \cite{bene2002}. Thanks to
the high spectral resolution of SWAS, the line is resolved and the measured line
width of few tens of \kms\, testified for the shock origin of the fundamental
water line. Assuming a common shock origin for the observed water in outflows,
SWAS and ISO give consistent results if we take into account that the
fundamental o-H$_2$O transitions can have a contribution from a cooler gas
component to which SWAS is sensitive but which is not traced by ISO.

\begin{figure}
  \includegraphics[width=15cm]{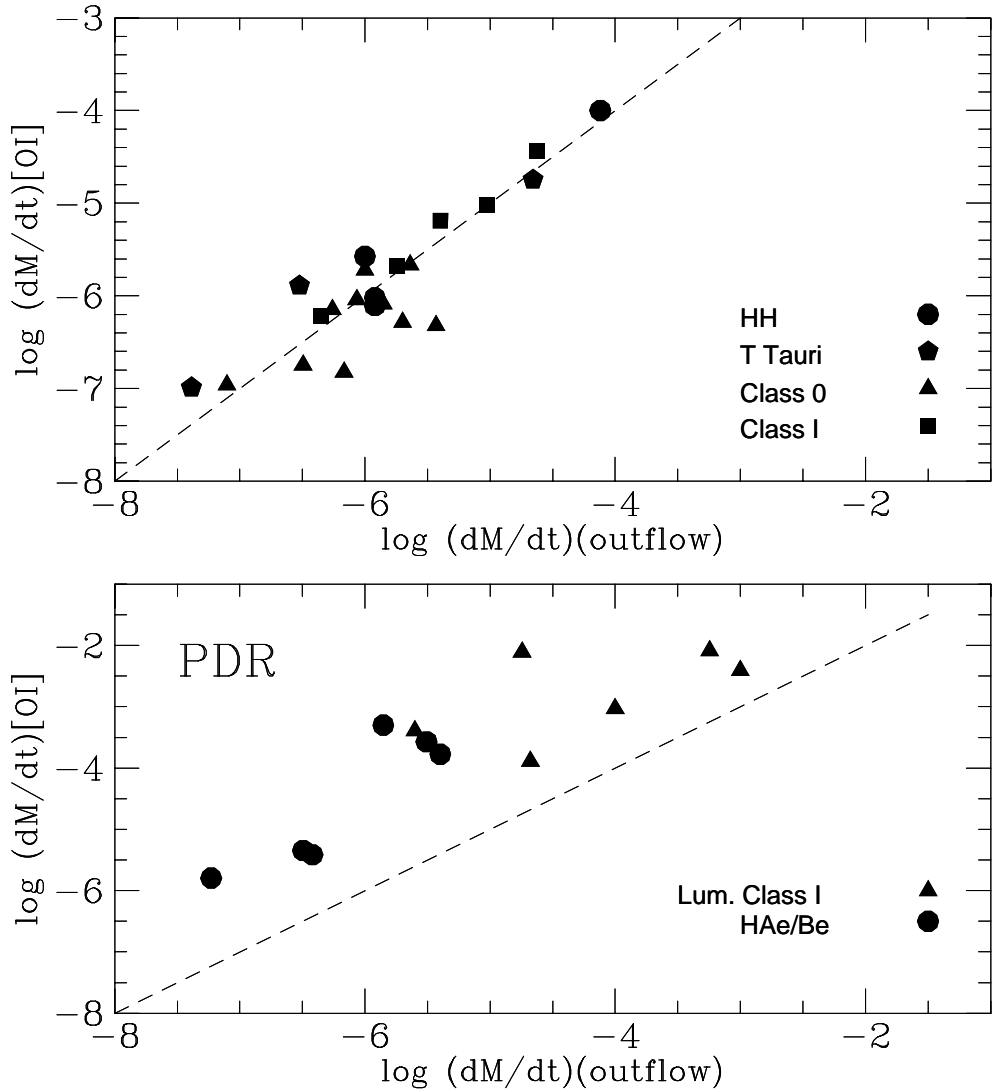}
  \caption{Mass loss rate determination from the [OI]63\um\, line vs that
derived from the CO molecular outflow for Class 0, low luminosity
Class I, HH objects and T Tauri stars (upper panel). In the lower
panel the same plot is shown for Luminous Class I and Herbig AeBe
stars.}
  \label{mdot}
\end{figure}

The atomic fine structure lines [OI]63\um\, and [CII]158\um\, have been observed
in all the sources of the sample, whilst the [OI]145\um\, is present only in the
66\% of the sources. In principle, these kinds of lines could be emitted from
the Photodissociation Region (PDR) originated by the FUV field provided by the
central object. However, it has been  found that the [CII]157.6\um\, emission is
quite constant in the different pointings of the same region. This result can be
explained by attributing the observed emission to the averaged
galactic FUV field, with the exception of the Herbig AeBe stars where this line
is attributed to the stellar PDR. The [OI]63\um\, line can originate from either
the PDR or by dissociative J shocks. In this type of shock, where the velocity
and the temperature of the shock front reach very high values
(v$\sim$50 \kms, T$\sim$10$^3$--10$^4$ K) sufficient to produce the dissociation
of the molecules and the ionization of the gas, the [OI]63\um\, line
dominates the cooling, so that its luminosity is proportional to the mass flux
into the shock providing a direct measure of the mass-loss rate (\.M$_{wind}$)
in the stellar wind \cite{holl85}. This rate can also be derived from the
millimeter CO maps of the molecular outflow. Assuming momentum conservation in
the interaction between the wind and the molecular outflow, in  case of J shock
the two mass-loss rate determinations should be equal. In the considered sample
of PMS objects, Saraceno et al. \cite{sara1999} found that the
equivalence of the two determinations is valid for Class 0, low luminosity Class
I, Herbig-Haro objects and T Tauri stars while for Luminous Class I and Herbig
AeBe stars the observed [OI]63\um\, emission exceeds that predicted by J shock
models (see Fig.\ref{mdot}). In the latter case a second and dominant component
is required, due the presence of HII regions and PDRs. Indeed, as expected,
Herbig AeBe stars have a PDR and high luminosity Class I sources are embedded in
HII regions and PDRs. The correlation shown by these objects is explained by the
fact that in HII regions and PDRs the [OI]63\um\, luminosity is proportional to
the bolometric luminosity: in fact, for luminous young objects, L$_{bol}$ is
proportional to the outflow mass loss rate, while for Class 0 and low luminosity
Class I this proportionality does not exist \cite{paolo}.

\begin{figure}
  \includegraphics[width=15cm]{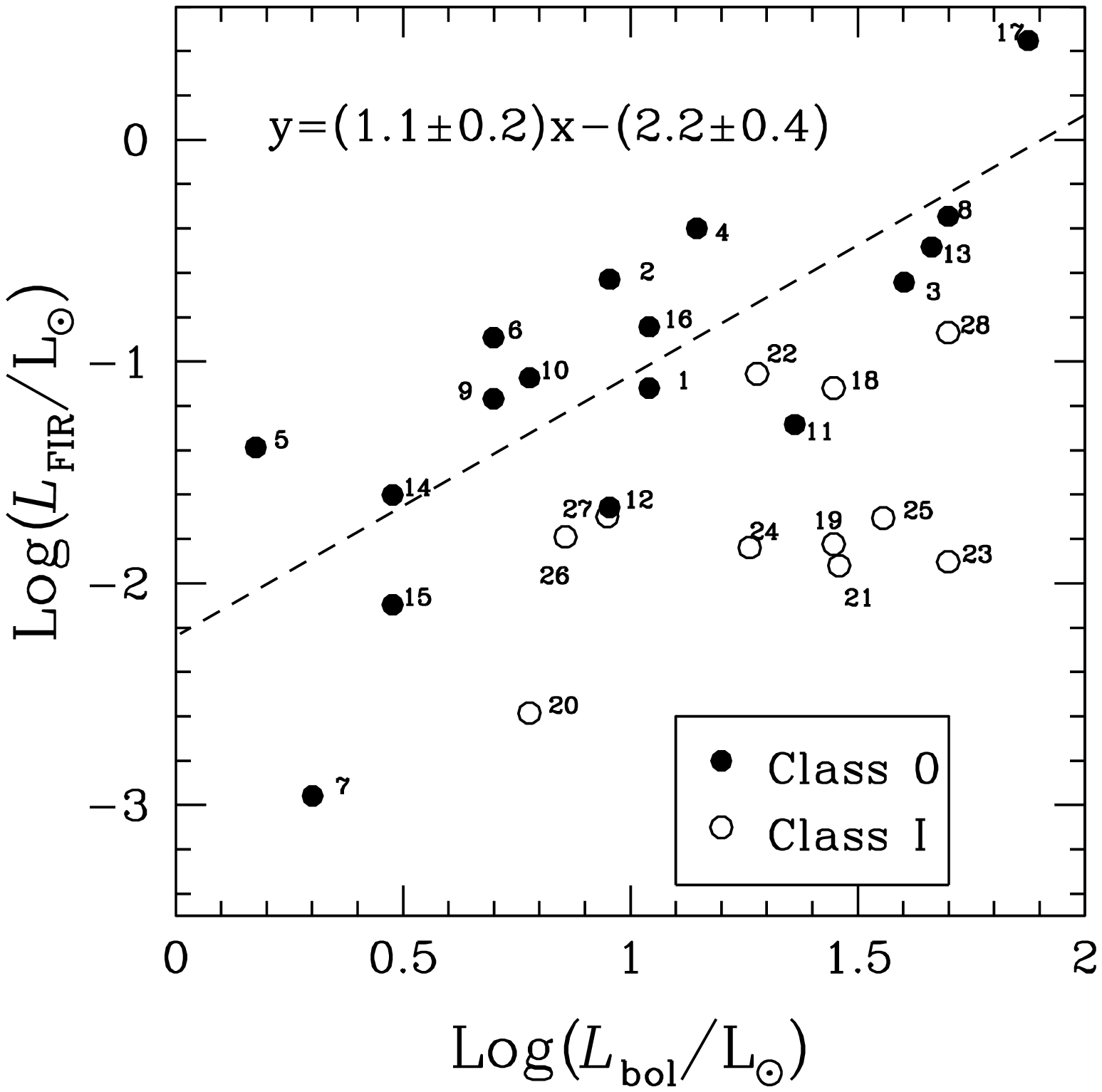}
  \caption{The FIR cooling vs bolometric luminosity for Class 0 and Class I
(filled and open circles, respectively) from \cite{nisi2002}. Numbers refer to
the source identification given in \cite{nisi2002}. The best fit through the
Class 0 data only is shown as a dashed line and its parameters are reported at
top of the figure.}
  \label{fir_bol}
\end{figure}

We can estimate the total FIR line cooling adding the contribution of all the
lines emitted in the LWS range, L$_{\rm FIR}$=L(OI)+L(CO)+L(H$_2$O)+L(OH). In
our sample L$_{\rm FIR}$ ranges from 0.01 to 0.1 \lsun\, and it is roughly equal
to the outflow kinetic luminosity as estimated from  the millimeter molecular
mapping. This circumstance demonstrates that the FIR cooling can be a valid
direct measure of the total power deposited in the outflow, not affected by
geometric or opacity problems like the determination of L$_{kin}$ or by
extinction problems like the near-infrared shocked H$_2$ emission. The total FIR
luminosity L$_{\rm FIR}$ results to be correlated with the bolometric luminosity
for the Class 0 sample with L$_{\rm FIR}$/L$_{\rm bol}\sim$10$^{-2}$ while for
Class I objects the L$_{\rm FIR}$ is lower of about one order of magnitude (see
Fig. \ref{fir_bol}). This difference is mainly due to the molecular emission
which is on average significantly larger in Class 0 than in Class I sources,
while the cooling due to atomic oxygen is fairly constant between the two
classes. This result is interpreted in terms of an evolution in the modality of
the interaction between the protostellar outflow and the circumstellar
environment, with an increasing influence of the progressively less shielded
interstellar FUV field \cite{nisi2002}. The strength of the outflow decreases
going from Class 0 to Class I and also the total gas cooling is expected to
decrease, but we observe a selective behaviour: the molecular cooling decreases
while the OI cooling does not. A possible explanation is that an inner jet
shock, responsible for most of the observed OI emission, decelerates the jet (or
collimated wind), while an outer ambient shock, which may have a bow morphology,
accelerates the ambient medium. The relative velocities of the two shocks depend
on the ratio between the wind and the ambient density. In Class 0 sources, the
ambient density is high and therefore the ambient shock velocity turns out to be
sufficiently low to produce strong non-dissociative C-shocks copiously emitting
in molecular transitions. In Class I sources most of the circumstellar matter
has already accreted onto the proto-stellar object so that the ambient density
may be sufficiently low to make the velocity of the ambient shock comparable to
the velocity of the wind. In this situation also the ambient shock can be
dissociative and cools down preferentially in OI emission, while some molecular
emission can arise from oblique shocks in the wing of the bow. A similar
mechanism can explain both the decrease of the molecular line cooling and the
almost constant behavior of the OI luminosity despite the expected decline of
outflow power. This trend is present also in Class II objects. In fact, in the
Herbig Ae/Be stars the molecular emission is observed in rare cases and
always only in the form of CO and OH \cite{gian1999}, indicating the dominant
role of the unshielded FUV field from the central object. The FUV field is also
responsible for a PDR where the observed atomic lines are excited
\cite{loren1999}. The small number of T Tauri stars (low-mass Class II objects)
studied is not enough to be representative of this class, making difficult to
draw any significant conclusion on the FIR spectra of the class. It is
remarkable that while few objects fit the above picture, since they show
only the atomic [OI] and [CII] lines, the prototypical T Tauri star, i.e. T Tau
itself, seems to contradict it, as it has a LWS spectrum extremely rich in
molecular transitions \cite{spino}. Such a spectrum can be however interpreted
as due to a tight interaction in the binary system composed by a young embedded
source, which may give rise to the molecular spectrum, and the optical PMS star,
whose UV field excites the observed [OI] and OH emission.

The reported results highlight the potentiality of the FIR spectra to
provide diagnostic indicators of the evolutionary status of young embedded
objects, allowing identifying specific criteria for their classification.

\subsection{LUMINOUS CLASS I}
The Luminous Class I sources are characterized by a high infrared
luminosity (L$_{\rm IRAS}>$10$^4$ \lsun) and are associated with
luminous outflows. Among the twelve sources of the sample (see Table
1), ten of them have LWS spectra dominated by
fine structure atomic emission and have been recently presented
and modeled by Montinaro et al. \cite{monti}. The other two sources are
NGC2024, a complex region which contains both a molecular
cloud core with strong molecular emission and an HII region \cite{gian2000} and
GGD27 IRS which is the exciting source of a powerful, collimated radio jet and
it is associated to the extremely high excited Herbig-Haro objects HH80 and 81
\cite{moli01}. These latter two sources are not discussed here.

\begin{figure}
  \includegraphics[width=15cm]{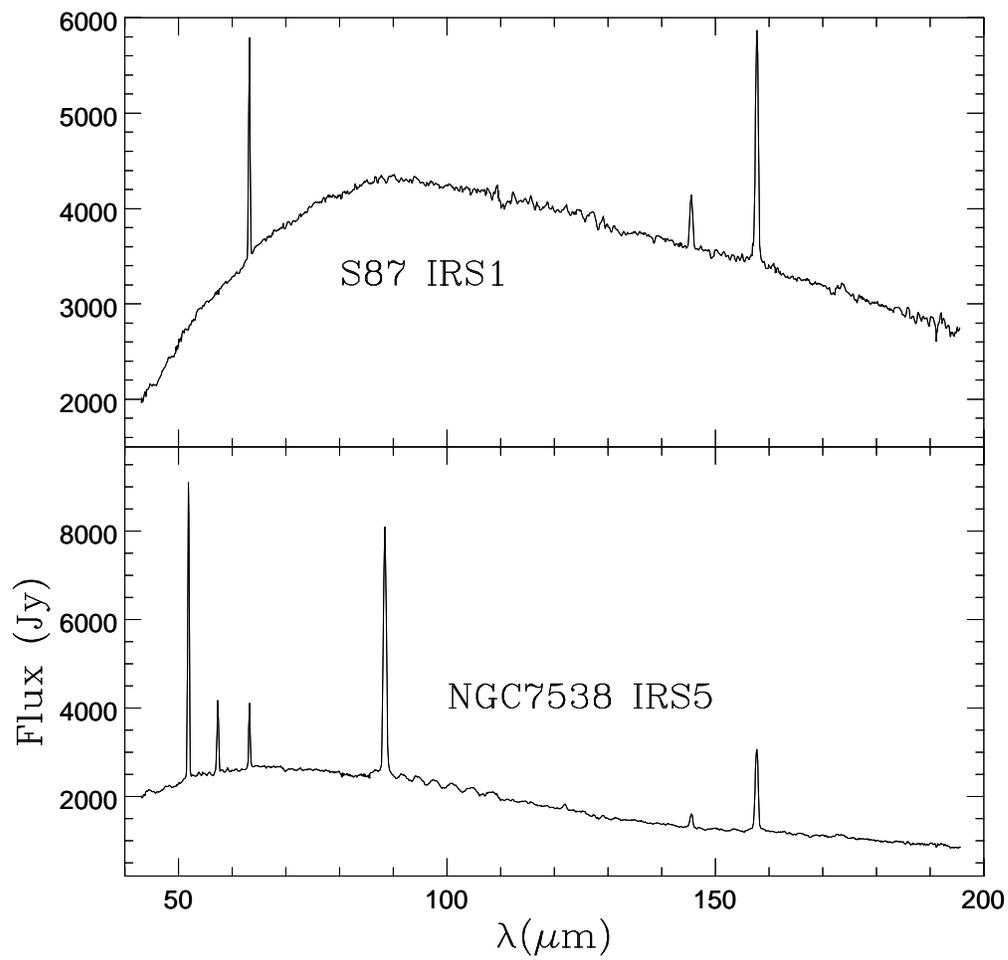}
  \caption{The LWS spectra of two Luminous Class I sources: S87 IRS1 (top panel)
and NGC7538  IRS5 (bottom panel).}
  \label{highlum}
\end{figure}

The remaining ten sources of our sample can be divided in two subgroups
depending on their emission line spectrum. In Figure \ref{highlum} we show the
LWS spectrum of two objects representative of the two subgroups. On average, the
less luminous ones have spectra typical of PDRs dominated by strong [OI] and
[CII] lines, while the more luminous ones, besides the [OI] and [CII] lines,
show also lines from intermediate ionized atoms [OIII], [NIII] and [NII] typical
of HII regions, photoionized by the stellar continuum.

The emission lines of these latter sources (NGC281W, W28A2, M8E, NGC7538 IRS1
and NGC7538 IRS5) can be accounted for by constant density
photoionization models. The higher ionization lines are generally
fairly well reproduced by the models, while the [OI] and [CII] emission lines
are in most cases underestimated by about an order of magnitude. In fact,
photoionization models fail to reproduce the PDR component that
is present at the interface between the HII region and the
molecular cloud. Using a PDR model to derive the size of the
hypothetical PDR that could account for the observed excess
emission, it is found that for most cases, except W28A2 and
NGC7538 IRS5, the needed PDR should have a size ranging from 0.03
to 1.0 pc, consistent with the observations.

The possibility that the [OI]63\um\, excess emission could
originate from J shocks, likely present in high luminosity
outflow sources, has been examined. W28A2 is the only source for which
the shock excitation can be the dominant process providing an
[OI]63\um\, emission comparable to the observed value.

In the other half of the sample of Luminous Class I
sources, namely AFGL437W, AFGL490, NGC6334I, S87 and NGC7538
IRS9, the only atomic fine structure lines observed are the [OI]
and [CII] lines. For these sources, which are classified as PDR,
the two line ratios [OI]63\um/[CII]158\um\,  and
[OI]63\um/145\um\, show that the density and the strength
of the FUV ionizing field \cite{Kauf99} cover very wide ranges (the
density span from about 15 to 10$^4$ cm$^{-3}$, while the field ranges from
about 6 to 10$^4$, in units of the Habing diffuse interstellar field).

From these observations, one may conclude that the FIR spectra of
high luminosity Class I sources follow an evolutionary scenario
from the PDR, dominated by OI and CII emission, originated in a
low energy field from the protostellar object, to the HII regions
where the intermediate ionization lines (OIII, NIII and NII)
become visible in the FIR, testifying that the FUV field of the
protostar is able to penetrate the dusty envelope where the source is
embedded.

\section{THE PROSPECTS OF THE HERSCHEL SATELLITE}
The next far infrared-submillimeter ESA mission will be the Herschel satellite
that will be launched in 2007. The satellite will
have a 3.5 m telescope and three instruments at the focal plane:
\begin{itemize}
\item Photodetector Array Camera and Spectrometer (PACS): equipped with
bolometer arrays for imaging photometry and photoconductor arrays for
spectroscopy. In photometry mode it simultaneously images two bands, 60-90 or
90-130 \um\, and 130-210 \um, over a field of view of 1.75$\times$3.5
arcmin$^2$. In spectroscopy mode, it images a field of view of 50$\times$50
arcsec$^2$, resolved on 5$\times$5 pixels corresponding to a spatial resolution
of $\sim$9.4 arcsec, with a spectral resolution of $\sim$175 \kms. For a detail
description of the instrument see \cite{pogl}.
\item Spectral and Photometric Imaging Receiver (SPIRE): a bolometer imaging
camera and a Fourier Transform Spectrometer (FTS). The photometer has a field of
view of 4$\times$8 arcim$^2$ and observes simultaneously at 250, 350 and 500
\um. The FTS has a field of view of 2.6$\times$2.6 arcmin$^2$ and a adjustable
spectral resolution of 300-15000 \kms\, at 250 \um. For a detail description of
the instrument see \cite{grif}.
\item Heterodyne Instrument for the Far-Infrared (HIFI): a heterodyne
spectrometer covering the range 157-221 and 240-625 \um\, with a maximum
spectral resolution of 0.06 \kms. The spatial resolution ranges from 12 to 46
arcsec. For a detail description of the instrument see \cite{graau}.
\end{itemize}

The study of the early phases of star formation is one of the key scientific
objectives of the Herschel mission. The better spectral and spatial resolution
together with the improved sensitivity of the instruments onboard Herschel with
respects to the ISO ones, will allow to make relevant progress in many
scientific problems highlighted by the ISO results. ISO has shown that the FIR
spectrum is a powerful tool to derive the evolutionary status of the PMS objects
and the FIR range is where most of the energy of the younger embedded objects is
emitted. However, the poor spatial and spectral resolution of the LWS has
allowed to address only the most prominent properties averaged on large spatial
regions, leaving open key questions that will be investigated by Herschel.

For example, HIFI will measure the line profiles up to 0.06 \kms\, allowing to
clarify the disputed origin, if envelope or outflow, of the molecular lines
observed with LWS in PMS objects. Moreover its spectral range will allow to
observe many other useful molecules as for example the S-bearing species that
are
good tracers of the shocks. PACS, with its line mapping capability and its
spatial resolution will be able to efficiently map the outflow associated with
the PMS objects, allowing to disentangle the different gas components and the
contribution of the different types of shocks (J-type/C-type). This kind of
observations will shed light on the complex structure of the outflow regions,
helping to constrain their physical conditions and to study the interaction
between the outflow and the surrounding molecular cloud.

The large field imaging capability of the Herschel cameras is suitable for
mapping
the large molecular clouds where the stars are forming. The typical distances
between protostars in the clusters range from $\sim$0.3 pc for low mass stars
\cite{gomez} to $<$0.04 pc for massive stars \cite{herbig}. This means that for
the intermediate and high mass stars observed with ISO, the FIR continuum
spectrum could be dominated by a younger companion present in the beam. The
spatial resolution of the Herschel cameras will greatly reduce this problem,
allowing
to derive the mass function of the clusters with a good sampling also for
the lower masses (the proto-brown dwarfs will be detectable). However
for the most crowded parts of the clusters the factor that will limits the
detection will be more likely confusion rather than sensitivity.

\end{document}